\documentclass[
a4paper, aps, prb, twocolumn, superscriptaddress, preprintnumbers, amsmath, amssymb, accepted=2023-09-07
]{quantumarticle}
\pdfoutput=1

\usepackage[T1]{fontenc}

\usepackage{graphicx}
\usepackage{dcolumn}
\usepackage{bm}
\usepackage{hyperref}
\usepackage{xcolor}
\usepackage[english]{babel}
\usepackage{braket}

\begin{document}

\selectlanguage{english}

\preprint{RBI-ThPhys-2023-21}

\title{Random Unitaries, Robustness, and Complexity of Entanglement}

\author{J. Odavi\'{c}}\email{jodavic@irb.hr}\affiliation{Ru\dj er Bo\v{s}kovi\'{c} Institute, Bijenička cesta 54, 10000 Zagreb, Croatia
}

\author{G. Torre}\email{gianpaolo.torre@irb.hr}
\affiliation{Ru\dj er Bo\v{s}kovi\'{c} Institute, Bijenička cesta 54, 10000 Zagreb, Croatia
}

\author{N. Miji\'{c}}\email{nenad.mijic@irb.hr}\affiliation{Ru\dj er Bo\v{s}kovi\'{c} Institute, Bijenička cesta 54, 10000 Zagreb, Croatia
}

\author{D. Davidovi\'{c}}\email{davor.davidovic@irb.hr}\affiliation{Ru\dj er Bo\v{s}kovi\'{c} Institute, Bijenička cesta 54, 10000 Zagreb, Croatia
} 
 
\author{F. Franchini}\email{fabio.franchini@irb.hr}\affiliation{Ru\dj er Bo\v{s}kovi\'{c} Institute, Bijenička cesta 54, 10000 Zagreb, Croatia
}

\author{S. M. Giampaolo}\email{salvatore.marco.giampaolo@irb.hr}\affiliation{Ru\dj er Bo\v{s}kovi\'{c} Institute, Bijenička cesta 54, 10000 Zagreb, Croatia
}

\date{\today}

\begin{abstract}
It is widely accepted that the dynamic of entanglement in the presence of a generic circuit can be predicted by the knowledge of the statistical properties of the entanglement spectrum. 
We tested this assumption by applying a Metropolis-like entanglement cooling algorithm generated by different sets of local gates, on states sharing the same statistic. 
We employ the ground states of a unique model, namely the one-dimensional Ising chain with a transverse field, but belonging to different macroscopic phases such as the paramagnetic, the magnetically ordered, and the topological frustrated ones.
Quite surprisingly, we observe that the entanglement dynamics are strongly dependent not just on the different sets of gates but also on the phase, indicating that different phases can possess different types of entanglement (which we characterize as purely local, GHZ-like, and W-state-like) with different degree of resilience against the cooling process.
Moreover, in some circumstances, we observe a {\it scrambling} effect by the algorithm that produces a Wigner-Dyson entanglement spectrum statistics on a state that does not obey a volume law for the entanglement entropy.
Our work highlights the fact that the knowledge of the entanglement spectrum alone is not sufficient to determine its dynamics, thereby demonstrating its incompleteness as a characterization tool.
Moreover, it shows a subtle interplay between locality and non-local constraints. 
\end{abstract}


\maketitle

\section{\label{sec:introduction} Introduction}

Entanglement is the most distinctive mark of quantum mechanics~\cite{Einstein1935, Bell1964} and an essential resource for many technological devices currently in development~\cite{NielsenBook, Degen2017, Ladd2010}. 
Therefore, it is easy to understand the growing interest in characterizing entanglement, especially in quantum many-body systems, as they represent the platform on top of which such devices are to be implemented. 
Characterization of entanglement is paramount in this context, as it is not just the sheer amount of entanglement which plays a crucial role in quantum applications. 
This is because some entangled states can be described efficiently also through classical resources. 
To provide an example, it is known that quantum circuits starting from factorized states diagonal in the computational basis and made by gates from Clifford's group~\cite{Gottesman1998}, can be efficiently simulated on a classical computer despite the amount of entanglement of the output state~\cite{Bravyi2016, Leone2021}. 
The resulting states are known as ``{\it stabilizer states}''.
However, adding gates outside Clifford's group, such as $T$-gates, makes it impossible to simulate the circuits efficiently on a classic computer~\cite{Chamon2014, Shaffer2014, Hinsche2022}. 
The difference between the two cases can be characterized by looking at the statistical properties of the entanglement spectra.
While in the second case, the output generally develops a Wigner-Dyson distribution in the entanglement spacing statistics, in the first one, the Poisson distribution is always obtained~\cite{Chamon2014, Shaffer2014, Zhou2020}. 
This difference plays a key role in the theory of quantum information, since only circuits doped with $T$-gates are capable of universal computation, as they can reach any state in the Hilbert space independently of the initial state~\cite{DiVincenzo2000, True2022}.

An alternative way of analyzing this difference is to use the concept of stochastic irreversibility (or robustness) of entanglement.
The idea behind this approach is to obtain information on the entanglement structure of a state by observing its evolution under the action of an entanglement cooling algorithm.
At its core, this algorithm is the usual Metropolis Monte Carlo protocol, with the cost function played not by the energy but by an entanglement measure. 
A transformation, chosen from a predefined set, is applied to the initial state, and the entanglement value of the new state is determined. 
The new state is accepted (retained) if the cost function has decreased and with a certain probability otherwise.
While states with a Poissonian statistic of the entanglement spectrum are not very resistant to this approach, i.e. after a few steps, the total amount of entanglement in the system tends to vanish, the cooling algorithm proves to be almost ineffective in states where this statistic follows a Wigner-Dyson distribution ~\cite{Chamon2014, Shaffer2014}. 
The different robustness of entanglement against a cooling algorithm has been linked to a concept of ``{\it complexity}'', coherently with the usual picture that a Wigner-Dyson distribution, stemming from the existence of strong correlations between the entanglement eigenvalues, indicate a higher complexity as well as higher robustness~\cite{Yang2017, True2022}.
This difference in robustness has proved to be extremely useful for distinguishing between the different dynamical phases present in quantum many-body systems~\cite{Yang2017} and falls within the recent interest of the quantum many-body community in random quantum circuits~\cite{Fisher2022,EisertComplex}.

In the present paper, we test how strong is the relationship between stochastic irreversibility and the entanglement spectrum statistic, by placing a particular emphasis on the choice of the initial states. 
In previous works (e.g. see  Ref.~\cite{Yang2017}), the states to which the cooling algorithm was applied were randomly generated. On the contrary, our starting states are the ground states of the one-dimensional quantum Ising model in its different macroscopic phases. 
As we will show, the entanglement spectrum of all these states follows a Poisson distribution for the level-spacing statistics, once some degeneracies are properly dealt with.	
Therefore, the action of the cooling algorithm could be expected to be independent of the initial state phase.
Instead, quite surprisingly, our results show a very different picture.

The most peculiar behavior we observe is associated with the ground states of models in a topologically frustrated phase~\cite{Dong2016, Maric2020, Maric2021, Maric2022, Torre2021, Torre2022}. 
They are obtained by imposing frustrated boundary conditions (that is, periodic boundary conditions with an odd number of sites) on systems with antiferromagnetic interactions.
The resulting ground states can be largely characterized as a linear superposition of single-dressed kink states~\cite{Giampaolo2019, Maric2022_bis} (topological solitons) and this representation allows us to describe, both qualitatively and quantitatively, the behavior of various physical quantities even in the presence of integrability breaking terms~\cite{Giampaolo2019, Maric2022_bis, Catalano2022, Maric2020_bis}.
In these states, after the cooling algorithm, we observe a stabilization of the entanglement to a reduced baseline value (finite, but strongly dependent on the size of the system), and any further reduction appears to be statistically unlikely. 

In the present work, we consider two sets of operations within the cooling algorithm. 
They are made of one- and two-body gates, with the latter acting only on neighboring spins.
In this sense, our algorithm can be considered made of local gates.
In the first set, we include only operations preserving the parity symmetry of the Hamiltonian which, as a consequence, cannot explore every state in the Hilbert space, thus generating a violation of ergodicity. 
On the contrary, in the second, the set of operations is extended to a complete set to ensure, at least in principle, access to any state.

While the end values of the entanglement obtained by the cooling algorithm starting from states with topological frustration are not qualitatively influenced by the choice of employed gates, this is not true for the ground states of the paramagnetic phase.
On them, the action of the cooling algorithm is practically negligible if the set of operations is limited to the one in which elements commute with the parity of the Hamiltonian, while their entanglement is quickly reduced if the larger set of gates is considered. 
This dependence on the gate set also allows distinguishing the paramagnetic phase from the ferromagnetic one, since on the latter the cooling algorithm is unable to destroy most of their entanglement, regardless of the gate set taken into consideration.

The difference between the two cases lies in the presence of entanglement of purely local nature in the paramagnetic case, while both the ground states in the ordered and topologically frustrated phases have long-range quantum correlations~\cite{Hamma2016, Giampaolo2019}.
Long-range entanglement is less affected by the action of local gates, and even less so as the system size increases.
On the other side, when the entanglement is local, local gates can easily reduce it, although the impossibility of exploring the whole Hilbert space with a reduced set of gates can still prevent its complete suppression. 

As noted above, the entanglement spectrum statistics of the initial states are always (mostly) Poissonian, and indeed this is also the case at the end of the cooling if the non-universal gate set is employed. 
However, when the universal one is used and when there is sufficient local entanglement to act on, the final states display a Wigner-Dyson (WD) entanglement spectrum statistics. 
Before our work, WD statistics has always been observed in states with volume law entanglement, but indeed we can confirm that this is not the case for our states.

The paper is organized as follows: In Sect.~\ref{sec:model} we introduced the model used to generate the input states of our cooling algorithm. 
Then, in Sect.~\ref{sec:Method} we describe in detail the cooling algorithm and the different sets of local gates. 
Afterward, in Sect.~\ref{sec:results}, we describe the results obtained, with a particular focus on how the size of the system, its quantum phase, and the gate set affects the evolution of the different states under the action of the cooling algorithm. 
In Sect.~\ref{sec:conclusion} we draw our conclusions.

\section{\label{sec:model}The Model}

As highlighted in the introduction, our goal is to apply an entanglement cooling algorithm (see also Sect.~\ref{sec:Method}) to states that are ground states of the same Hamiltonian but in different macroscopic phases. 
We focus on the ground states of the one-dimensional spin-1/2 transverse field Ising model (TFIM), since it is a prototypical example of a many-body system possessing different macroscopic phases.
It is defined by the following Hamiltonian, 
\begin{equation}
    H = J \sum\limits_{l=1}^{N} \sigma_{l}^{x} \sigma_{l+1}^{x} - h \sum\limits_{l = 1}^{N} \sigma_{l}^{z},  \label{HamDef}
\end{equation}
where the $\sigma^{\alpha}$ with $\alpha=x,\, y, \,z$ are the Pauli operators. 
Limiting our analysis to systems with periodic boundary conditions ($\sigma^{\alpha}_{N+1} = \sigma^{\alpha}_{1}$) made by an odd number of spins ($N = 2M  + 1$ with $M \in \mathbb{N}$), the so-called frustrated boundary conditions (FBC), the model admits three distinct phases.
When the local field dominates over the interaction term between spins, the system is in the paramagnetic phase (PARA), characterized by a gapped spectrum and a vanishing spontaneous magnetization in the longitudinal directions.
On the contrary, when the interaction term dominates over the local field ($|J|>|h|$), we can realize two inequivalent phases depending on the sign of the interaction $J$. 
If the interaction is ferromagnetic (FM), i.e. when it favors parallel alignments ($J<0$), the system shows a gapped magnetically ordered phase, with a finite magnetization in the $x$- direction~\cite{FranchiniBook}. 
On the other hand, in the case of an antiferromagnetic interaction (AFM, $J>0$) the system is in a gapless topological frustrated phase~\cite{Dong2016, Maric2020, Maric2021, Maric2022, Torre2021, Torre2022}, where the spontaneous magnetization is destroyed by the presence of a delocalized kink excitation.
We remark that the TFIM is also integrable. 
We do not use this feature in our analysis, and we do not expect it to influence our results: to check this assumption, in Sec.~\ref{sec:nonint} we will also add an integrability breaking term and apply the same algorithm to the resulting ground state.

\begin{figure}[t]
	\centering
\includegraphics[width=\columnwidth]{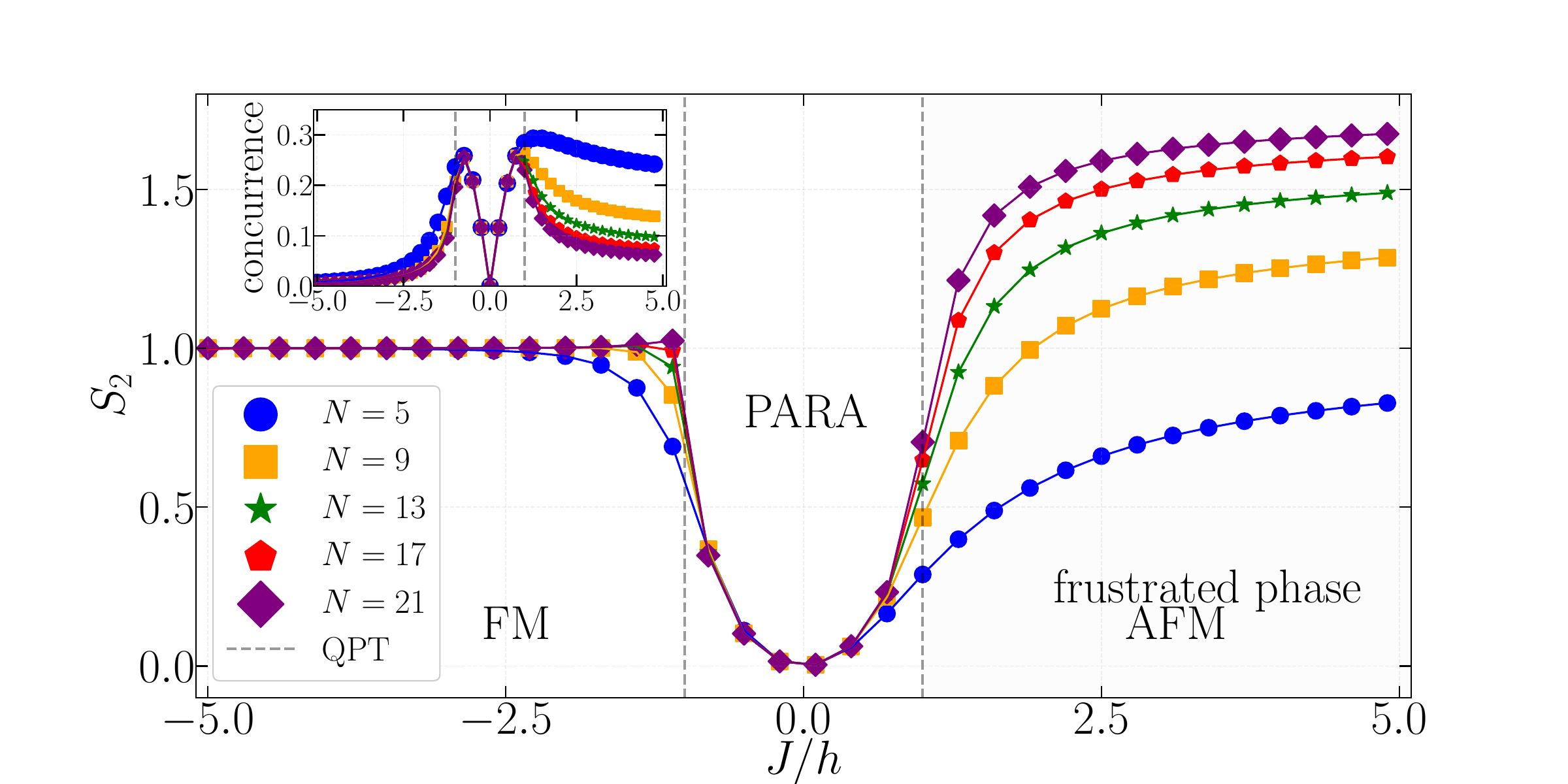}\caption{Half-chain R\'{e}nyi-2 entropy 
of the transverse-field Ising model. 
The quantum phase transitions 
(black dashed lines) separate the paramagnetic phase ($\vert  J/h  \vert \! < \! 1$) from the ferromagnetic ($ J/h \!< \! -1$) and the frustrated antiferromagnetic one ($ J/h \!> \!1$). 
The excess entanglement in the frustrated AFM regime is an increasing function of the system size, which saturates at the thermodynamic limit~\cite{Giampaolo2019}. In the inset, we plot the 
nearest-neighbor concurrence~\cite{Wootters1998}, where the excess concurrence in the frustrated phase decreases with the chain length. In the thermodynamic limit, the curves would be symmetric around the origin. } 
	\label{PhasesEnt}
\end{figure}

A way to discriminate among the three model's phases is by looking at the various kinds of bipartite entanglement~\cite{NielsenBook, Amico2008} that can be quantified by the R\'{e}nyi entropies, defined as,
\begin{align}
    S_{\alpha} (\rho_{\rm A}) = \dfrac{1}{1 - \alpha} \log_2{{\rm Tr} \left[ \rho^{\alpha}_{\rm A}  \right]}, \label{entropydef}
\end{align}
which depends on the parameter $\alpha  \in [ 0,1) \cup (1,\infty ]$. 
In eq.~\eqref{entropydef}, $\rho_{\rm A } \equiv {\rm Tr}_{\rm B} \ket{ \Psi}\bra{\Psi } $ is the reduced density matrix obtained by tracing out from the ground state $\ket{\Psi}$ all the degrees of freedom of spins that lie outside the subset $A$.
In the limit $\alpha \to 1^+$, the R\'{e}nyi entropies reduce to the von Neumann entropy~\cite{NielsenBook}: $S_{1} (\rho_{\rm A}) = - \sum_{k} \lambda_{k} \log_2{ \lambda_{k}}$, where $\{ \lambda_{k} \}$ 
is the set of eigenvalues of the reduced density matrix $\rho_{\rm A}$. 

In Fig.~\ref{PhasesEnt} we present the behavior of $S_{2} (\rho_{\rm A}) $ as a function of the ratio $J/h$ for $h>0$ in the case in which $A$ is made by $(N-1)/2$ contiguous spins.
For the sake of simplicity, we refer to the string made by $(N-1)/2$ contiguous spins as the \emph{half chain}. 
Differently from the other phases, in the topologically frustrated one, we observe a relevant dependence of the entanglement on the size of the chain that can be explained by taking into account that in such a phase the ground states are characterized by a delocalized excitation that increases the total amount of entanglement~\cite{Giampaolo2019}. 

While the half-chain R\'{e}nyi-$\alpha$ entanglement entropy captures the non-local nature of entanglement, the concurrence~\cite{Wootters1998} measures its local contribution.
It is defined starting from the reduced density matrix obtained by tracing out the degrees of freedom of every site of the chain but two, which in this case we take as neighboring sites. 
In the inset of Fig.~\ref{PhasesEnt}, we observe that contrary to what happens to $S_2$, the contribution of the delocalized excitation in the frustrated phase decreases as the chain length increases and vanishes in the thermodynamic limit.
As a consequence, for large systems, the concurrences of the FM and the AFM phases coincide.
However, due to the complexity of the cooling algorithm, our system will always be well below such a limit.
Hence, for the analyzed cases, the amount of local entanglement of a ground state in the presence of a topological frustrated system will be relatively higher than the one coming from the FM phase.

\begin{figure}[t!]
	\centering
\includegraphics[width=\columnwidth]{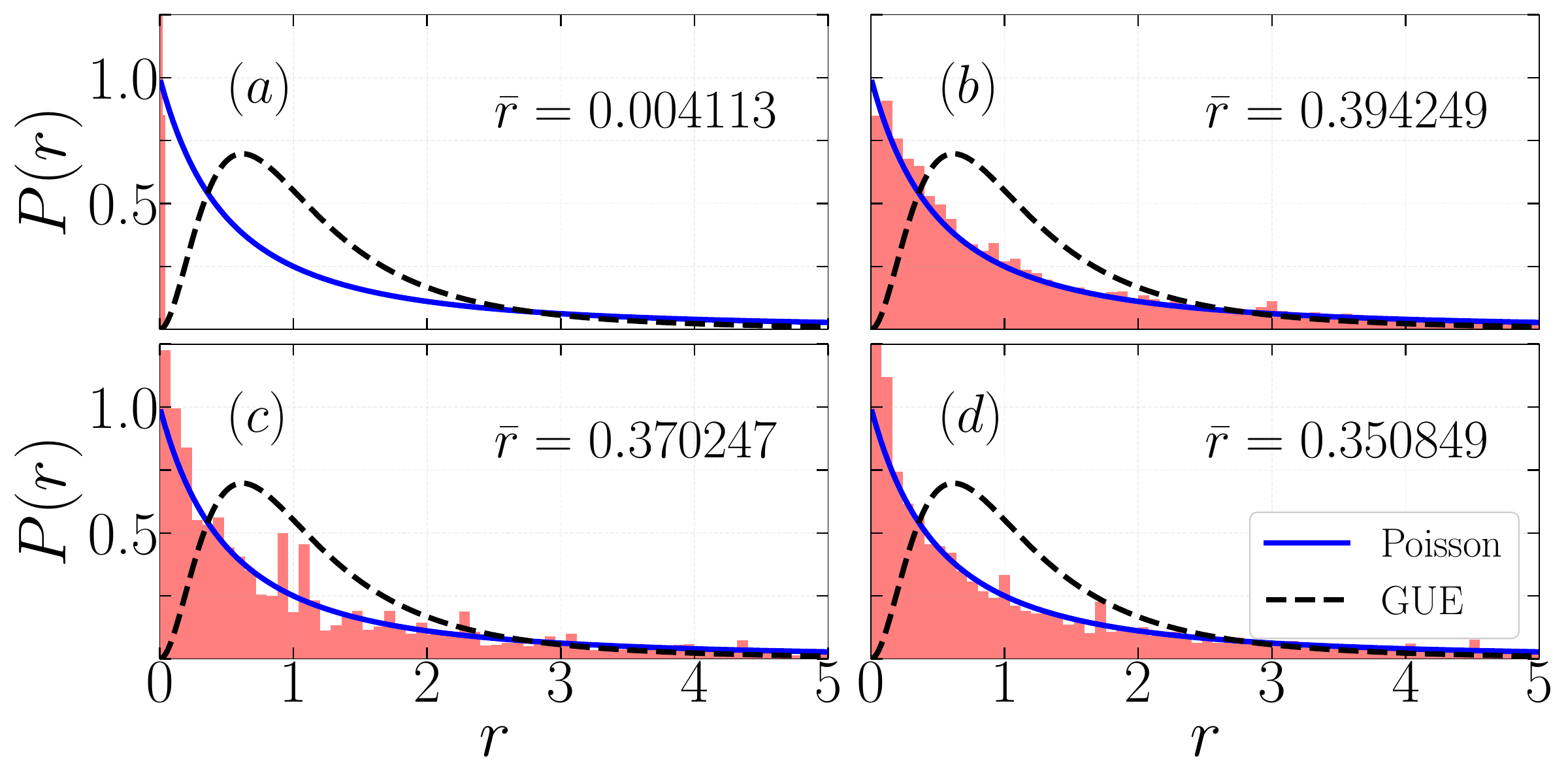}
\caption{ Histograms for consecutive entanglement spectrum spacing of the reduced density matrix (RDM) eigenvalues of the ground states in different regimes of the TFIM  Hamiltonian for $N \!=\!35$ spins. Number of bins used is $125$.  We superimposed the analytical curves for the Wigner-Dyson (dashed) and Poisson (continuous) distribution for comparison. 
Due to the large system/subsystem sizes, we obtain good statistics within each single shot realization, which is what is plotted. 
The different regimes taken into account are: \textit{(a)} Deep ferromagnetic (FM) regime with $J/h \! = \! -2.5$; 
\textit{(b)} Ferromagnetic (FM) regime close to the quantum phase transition point at $J/h \! = \! -1.05$; \textit{(c)} Paramagnetic (PARA) regime with $J/h \! = \! 0.75$; 
\textit{(d)} Frustrated antiferromagnetic (frus. AFM) regime with $J/h \! = \! 2.5$. 
Except for results in panel $(a)$, the entanglement spectrum spacing statistic always follows a Poisson distribution. 
In panel $(a)$, the majority of the spacing falls in a narrow bin close to the origin because of the degeneracies of the entanglement spectrum for subsystems much bigger than the correlation length. For more details on the computation, see Appendix~\ref{sec:Appendix2}.}
	\label{InitialES}
\end{figure}

Despite the differences in the entanglement entropies discriminating the various phases of the model, the statistical properties of the entanglement are almost the same. 
To highlight this fact, we focus on the probability density function of the consecutive level spacing ratio. 
Ordering the set of eigenvalues $\{ \lambda_{k} \}$  of the reduced density matrix $\rho_A$ from the smallest to the biggest, the set of the consecutive level spacing ratios are
\begin{equation}
    r_{k} = \frac{\lambda_{k + 1} - \lambda_{k}}{\lambda_{k} - \lambda_{k - 1} }, \quad k = 2, 3, 4, ..., 2^{ \lfloor N/2 \rfloor} - 1. \label{ratio_text}
\end{equation}  
The elements of this set are distributed accordingly with different statistics, see Appendix~\ref{sec:Appendix2}. The results obtained are shown in Fig.~\ref{InitialES} for four different parameter choices.
We observe a significant deviation only close to the classical ferromagnetic point. 
It is known \cite{Franchini2011} that in this phase for partitions much bigger than the correlation length the entanglement spectrum develops exact degeneracies, which explains the coalescing of so many spacing toward zero.
By manually removing the degenerate eigenvalues from the statistics, one can obtain a Poissonian distribution. 
Increasing the correlation length by moving closer to the quantum phase transition also removes the degeneracies and the plot shows a greater consistency with Poissonian curve. 
For each plot, we also report the spacing ratio, whose vicinity to $\tilde{r}_{\rm Poisson} \simeq 0.386$ indicates good Poissonian statistics (see Appendix~\ref{sec:Appendix2} for more details).

\section{\label{sec:Method} The Entanglement Cooling Algorithm }
 
Even if the Hamiltonian in eq.~\eqref{HamDef} is integrable, we recover the ground states in the different phases by exact diagonalization~\cite{Sandvik2010}. 
The reason is two-fold. 
First, with the explicit form in the spin basis (as compared to the form in terms of Bogoliubov fermions) the GS is more suitable for generic gate applications. 
A second reason for this choice comes from the computational bottleneck associated with the large number of gates applications needed to resolve the behavior, which already for the system sizes accessible by exact diagonalization requires several days of computation.

Once we have obtained the GS, we apply to it the entanglement cooling algorithm to test the stochastic irreversibility of its entanglement.
This algorithm is nothing else but a Metropolis Monte Carlo (MC) algorithm~\cite{BinderBook} in which each step is made of two successive operations. 
The first one is the application of a single randomly chosen gate to the state coming from the previous step, i.e. a unitary `time' evolution operator that transforms the state and can modify the value of the entanglement.
The gates can be written as \mbox{$U_l^{(k)}\!=\!\exp{(i O_{l}^{(k)} \Delta t)}$} with $\Delta t = \pi / 10$ and are applied at randomly selected neighboring pairs of spins $\{l,l+1 \}$. 
The different kinds of gates generator $O_l^{(k)}$ are selected with uniform probability from the sets of deterministic operators tabulated in Table~\ref{table1}. 
Concerning the Hamiltonian in eq.~\eqref{HamDef}, the two sets hold very different properties. The first is made by gates that 
preserve the parity symmetry of the Hamiltonian along $z$-direction. 
On the contrary, the gates in the second set, either violate the parity symmetry or are generated by non-integrable operators.
It is worth noting that, while both sets of gates alone do not constitute complete sets of operations able to provide universal quantum computation, the two sets together do (see Ref.~\cite{Barenco1995}).

\begin{table}[b!]
\begin{center}
{\small
\begin{tabular}{|c | c |} 
 \hline
 \textbf{Set 1} & \textbf{Set 2} \\ [0.5ex] 
 \hline
 $O^{(1)}_{l} = \sigma^{z}_{l} \otimes \mathbb{I}_{l + 1} + \mathbb{I}_{l} \otimes \sigma^{z}_{l + 1}$  & $ O^{(4)}_{l} = \sigma^{x}_{l} \otimes \mathbb{I}_{l + 1} + \mathbb{I}_{l} \otimes \sigma^{x}_{l + 1}$  \\ 
$O_{l}^{(2)} = \sigma^{x}_{l} \otimes \sigma^{x}_{l+1}$  & $O^{(5)}_{l} = \sigma^{y}_{l} \otimes \mathbb{I}_{l + 1} + \mathbb{I}_{l} \otimes \sigma^{y}_{l + 1}$  \\
 $O_{l}^{(3)} = \sigma_{l}^{y} \otimes \sigma_{l+1}^{y}$  & $O^{(6)}_{l} = \sigma^{z}_{l} \otimes \sigma_{l+1}^{z}$ \\
 \hline
\end{tabular}
\caption{\label{table1} Table of local deterministic gate generators. Set 1 is parity preserving. Sets 1 \& 2 taken together form the universal set~\cite{Barenco1995}. }}
\end{center}
\end{table}

The second operation is a filtration procedure, in which the new step can be accepted or rejected depending on the change of a cost function.
The main difference between our algorithm and the traditional Metropolis MC ones is that in these last the cost function is represented by the energy, while in our case is the half-chain entanglement.
We remark that this choice implements an interesting interplay between the locality of the applied gates and the global nature of the cost function.
The new state is accepted or rejected with a probability equal to $\min \{ 1, {\rm exp} ( - \Delta \bar{S}_{\alpha}/T ) \}$ where $\Delta \bar{S}_{\alpha} = \bar{S}^{\rm new}_{\alpha} - \bar{S}^{\rm old}_{\alpha}$ denotes the difference between the averaged R\'{e}nyi-$\alpha$ entropies ($\bar{S}_{\alpha} = N^{-1} \sum_{l} S_{\alpha,l}$) where the average is made over all sets made of $(N-1)/2$ contiguous spins. 
Even if in principle the average is taken over $N$ different partitions, at each step only two entropies need to be recomputed since local operations inside a single partition do not change the value of the entanglement.

\begin{figure}[t!] 
	\begin{center}
\includegraphics[width=0.3\textwidth]{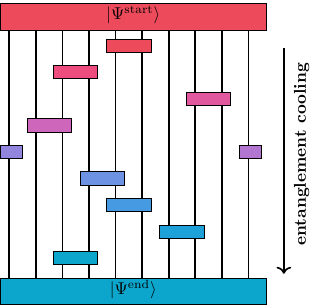}
\caption{\label{AlgorithmScheme} 
Cartoon of the entanglement cooling algorithm in terms of a quantum circuit. 
Provided with a quantum many-body state, $\ket{ \Psi^{\rm start}} $ random unitaries (colored and small rectangles) act 
on two neighboring sites that are represented by a pair of neighbor vertical lines~\cite{Evenbly2022} (Periodic boundary conditions are assumed). 
The resulting state can be accepted or rejected depending on the change in the averaged R\'enyi-2 entropy and the step-dependent pseudo-temperature.}
\end{center}
\end{figure}

\begin{figure*}[t]
	\centering
\includegraphics[width=18cm]{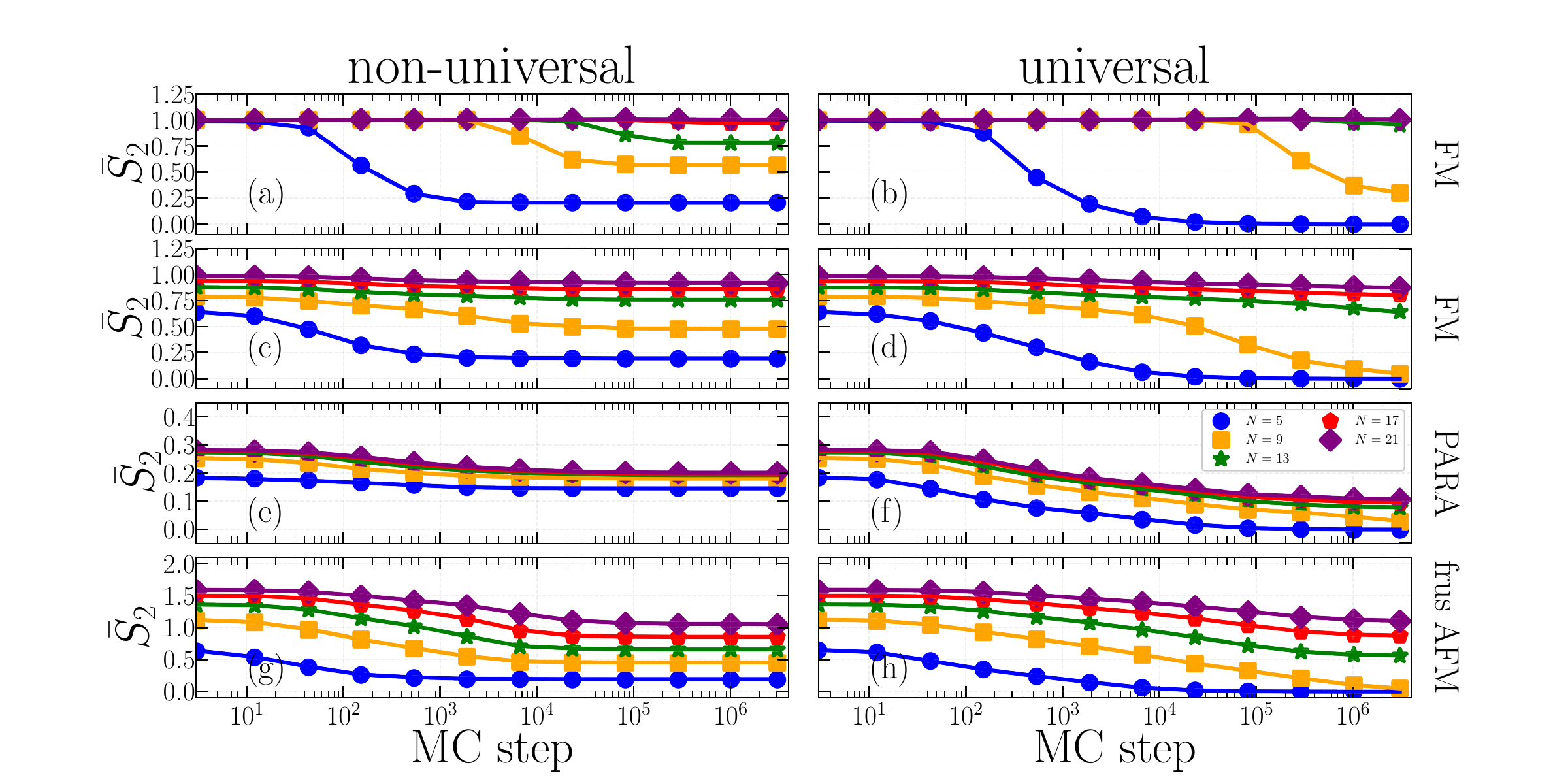}
\caption{ Averaged half-chain R\'{e}nyi-2 entropy during the entanglement cooling over $M = 96$ Metropolis MC trajectories for the ground states of the Hamiltonian given by eq.~\eqref{HamDef} in different macroscopic phases. The plots in the left columns are obtained by applying local gates only from the first set (non-universal), while in the right column, the cooling algorithm applies local gates coming from both sets (universal); see Table~\ref{table1}. The choice of points in the phase diagram is the same as in Fig.~\ref{InitialES}. In panels (\textit{a}) and (\textit{b}), the results are obtained for the deep ferromagnetic (FM) ground-state ($J/h \! = \! -2.5$). In the panels (\textit{c}) and (\textit{d}), we present the results in the ferromagnetic (FM) regime close to the quantum phase transition point ($J/h \! = \! -1.05$); see Fig.~\ref{PhasesEnt}. Next, in panels (\textit{e}) and (\textit{f}), the evolution obtained starting from ground states of the paramagnetic (PARA) phase ($J/h = 0.75$) is displayed. In the bottom panels (\textit{e}) and (\textit{f}), the evolutions obtained for ground states in the topologically frustrated antiferromagnetic (frus AFM) regime ($J/h = 2.5$) are shown.  }
\label{EntEvolution}
\end{figure*}

Using the von Neumann entropy ($\alpha = 1$) or any of the R\'{e}nyi entropies ($\alpha \neq 1$) the entropy cooling algorithm leads to similar behavior. 
In particular, we observe that statistically, the inequality $S_{q} \ge S_{q'}$ for $q < q'$ is satisfied at any point of the algorithm's execution~\cite{Muller2013}. 
For our analysis we employ the $\alpha\!=\!2$ R\'{e}niy entropy~\cite{Horodecki2002, Plenio2007, GiampaoloMontangero2013, NielsenBook}, 
since it can be computed faster than the von Neumann entropy since it can be computed directly through matrix-matrix multiplication, without the need for a diagonalization algorithm required by the von Neumann entropy. 
The main advantage of our approach is that it can be, and it has been implemented very efficiently (and in a parallelized way) using the architecture of modern graphical processing units (GPU) achieving a speed-up in matrix-matrix operation compared to standard CPUs (see Appendix~\ref{sec:Appendix1} and Ref.~\cite{Mijic2022}.)
Therefore, R\'{e}nyi-2 entropy is an efficient choice on top of being observable via atomic and optical experiments~\cite{Lesche2004, Bovino2006, Abanin2012, Islam2015, Kaufman2016, Brydges2019}. 
Moreover, the $S_{2}$ is also connected to the information scrambling in quantum systems with averaged out-of-time-ordered correlations (OTOC), that are used to diagnose the onset of quantum chaos~\cite{Hosur2016}.

The cooling part of the algorithm refers to a fictitious temperature $T$ decreasing as a function of the Metropolis MC steps. 
Indeed, the temperature is lowered in an evenly spaced logarithmic temperature grid spanning the range $T \!\in\! \left[ 10^{-4}, 10^{-8} \right]$. 
In the literature, this procedure is usually referred to as the simulated annealing method and is often used in the efficient search for global minimum in various high-dimensional problems~\cite{Shaffer2014}. 
Moreover, as is typical in these kinds of approaches, we perform an average of over $M$ different Metropolis MC trajectories or stochastic quantum circuits.

\section{\label{sec:results}Results}

Having described both the entanglement cooling algorithm and the model used to generate the input states of the algorithm, we move to illustrate our results.
The different types of evolutions of the entanglement obtained with the algorithm are summarized in Fig.~\ref{EntEvolution}. 
The plots in the left columns are obtained by choosing local gates uniformly at random from the first set, while in the right one, all local gates are considered to ensure a universal quantum computation (1 \& 2). 
For more information about our efficient computational implementation, allowing us to reach systems up to $N = 21$ spins with a considerable amount of Metropolis MC steps and statistics, see Appendix~\ref{sec:Appendix1}.

Deep in the FM cases, we observe that except for the smaller systems, the state displays almost perfect robustness of the entanglement regardless of the set of gates employed. 
This can easily be explained since in the FM phase, the ground state is well approximated by a $N$-qubit Greenberger–Horne–Zeilinger (GHZ) state~\cite{Greenberger1989}.  
Moving closer to the quantum phase transition adds local correlations on which the algorithm can act, but their contribution is hardly noticeable on the R\'enyi-2 for the parameters we chose. 
The GHZ state represents a global symmetric superposition of two orthogonal fully factorized states. 
Such states are well-known in the theory of quantum information, since they maximize the multipartite entanglement~\cite{Dur2000} while retaining no bipartite entanglement after local measurement on one of its parts~\cite{Coffman2000}. 
In such a state, all the quantum correlations come out from the peculiar global superposition~\cite{Hamma2016} and, therefore, it comes as no surprise that local gates acting at most on two neighboring sites of the chain are unable to change this entanglement. 
To simulate a global interaction with local interactions, we need a series of the latter which, taken individually, will increase the value of the entanglement of the state. 
The probability that this happens, considering how the cooling algorithm selects the different states, decreases exponentially as the size of the chain increases. 

\begin{figure}[t]
	\centering
\includegraphics[width=\columnwidth]{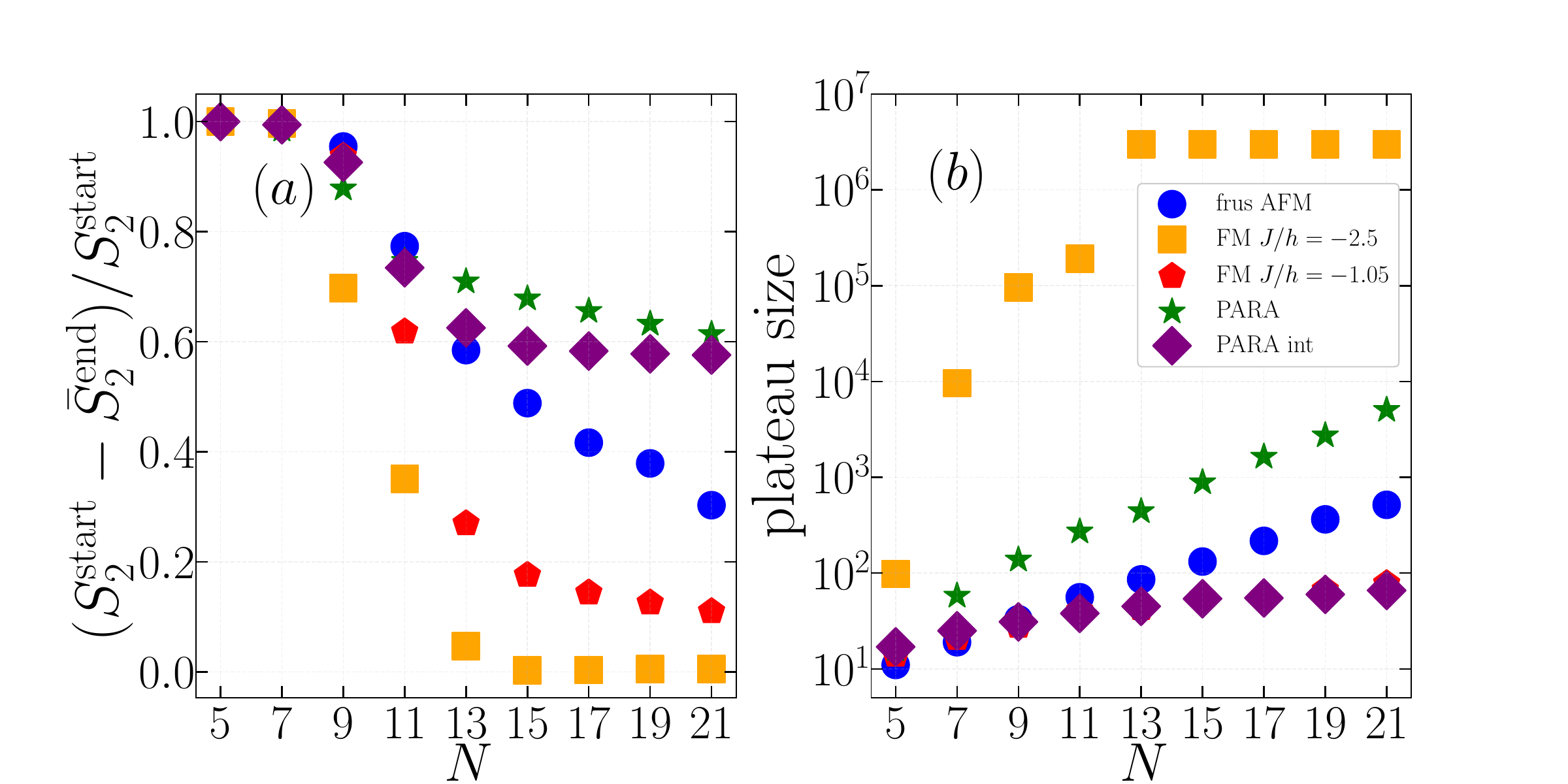}
\caption{Finite-size analysis of the plateaus reached when the entanglement cooling algorithm is applied using the universal gates set. 
Left $(a)$ panel: Relative difference for the various states between their half-chain R\'{e}nyi-2 entropy before and after the cooling, i.e. \textit{start} and \textit{end}, respectively. The end value of the entanglement is averaged over the different MC trajectory realizations. 
Right $(b)$ panel: Initial plateaus size, defined as the number of MC steps after which the entanglement deviates by more than 5\% from its starting value for the different cases considered in Fig~\ref{EntEvolution}.}
	\label{Plateaus}
\end{figure}

In the PARA regime, the dynamic of the entanglement is strongly affected by the chosen sets of local gates. 
After applying the first set, we observe that, after a small transient in which the total entanglement is slightly reduced, we arrive at an almost size-independent plateau for the entanglement. 
Conversely, when the complete set of local gates is taken into account the entanglement is considerably reduced, hence signaling a clear difference from the FM phase, where the entanglement cannot be removed.

In the final case, when the cooling algorithm is applied to the frustrated AFM ground state, both similarities and differences compared with the previous cases are evident. 
Similar to the PARA case, the entanglement is not completely robust, while akin to the FM phase the overall behavior of the entanglement evolution does not depend on the set of gates used in the algorithm. 
In particular, the entanglement remains initially locked to its initial value before being reduced.
However, this reduction does not proceed until the entanglement is completely removed, but it rather stops at a size-dependent baseline value. Moreover, the rate of the entanglement cooling scales with the system size $N$. 
Increasing the system size, the rate of decay becomes less steep, indicating that the entanglement becomes extensively robust, in agreement with the delocalized single-particle interpretation~\cite{Giampaolo2019}. 

In Fig.~\ref{Plateaus}, we present a finite-size analysis of the (average) values reached by the entanglement entropy at the end of the cooling algorithm. 
In all cases but the frustrated one it is clear that we have reached a plateau, which constitutes a more or less significant reduction of entanglement, although the difference from the starting value decreases with the system size. 
For the AFM frustrated phase, it is not possible to extrapolate from our data the asymptotic regime. 
Nevertheless, we expect that the final value will keep increasing algebraically with the system size towards a value that is only slightly reduced compared to the initial one.
This is consistent with the progressive difficulty of the algorithm in destroying the long-range entanglement in the W-state structure of the frustrated states as the number of sites is increased (while the local entanglement in the dressed kink states can be attacked and reduced)~\cite{Odavic2022}.
Interestingly, the initial plateau for which the entanglement remains within 5\% of its initial value seems to increase with the system size in all considered cases, as shown in the right panel of Fig.~\ref{Plateaus}.

Changing the cooling procedure, e.g. modifying the temperature gradient in terms of the MC steps, would yield slightly different results for the averaged final entanglement entropy.  
Nevertheless, the scaling of robustness with system size $N$ behavior looks ubiquitous and not influenced by these choices. 

Thus, though all considered initial states share largely the same initial statistic for the entanglement spectrum, we see very different behaviors under the cooling evolution.
This fact indicates that the reduced density matrix eigenvalue statistics are not sufficient to completely capture the underlying complexity of entanglement, in the sense of its robustness~\cite{Yang2017, True2022}. 
However, the various phenomenologies can be explained easily.
Let us start by considering the FM and PARA phases. 
Both phases present a finite energy gap that separates the ground states from the overlying excited states. 
Therefore, the ground states can be obtained starting from their relative classical points ($\lambda \! = \! 0 $ for the PARA phase and $ \lambda \! = \! - \infty $ for the FM phase) using a quasi-adiabatic continuation~\cite{Hastings2005}. 
This process generates entanglement of local nature that can be easily removed by an entanglement-cooling algorithm such as the one used in~\cite{Hamma2016}. 
However, the two classical points present a notable difference. In the case of the paramagnetic phase, the classical ground state is single, and therefore the local entanglement generated by the adiabatic continuation process is the only one present in the system.
In principle, one could therefore expect complete destruction of entanglement in the process. Our results indicate that a plateau at a finite value is reached also in this case. 
We speculate that this is due to the simulated annealing: after some local entanglement is removed by the algorithm, the systems reach an equilibrium between the instances in which moves that increase the entropy are accepted and those that are able to reduce the entanglement.
Conversely, at the classical ferromagnetic point, the ground state is doubly degenerate. 
Adding a small transverse field breaks the symmetry by selecting a particular global superposition, producing a state similar to a GHZ state~\cite{Maric2022_bis}. 
Therefore, the quasi-adiabatic deformation process can be seen starting from this state and
the local entanglement thus generated is added to that
associated with the global superposition.

Furthermore, by progressively increasing the magnetic field, one can follow a quasi-adiabatic deformation of the ground state that acquires additional local entanglement in addition to the non-local one associated with the global superposition.
While the local entanglement can be directly removed by the cooling algorithm, the global one is resilient against its action, since it is based on local gates, unless very small sizes are considered. 
Hence, the resilience of entanglement in the FM phase that we see in our data to a value close to $\log_2{(2)} = 1$.

\begin{figure}[t]
\centering
\includegraphics[width=\columnwidth]{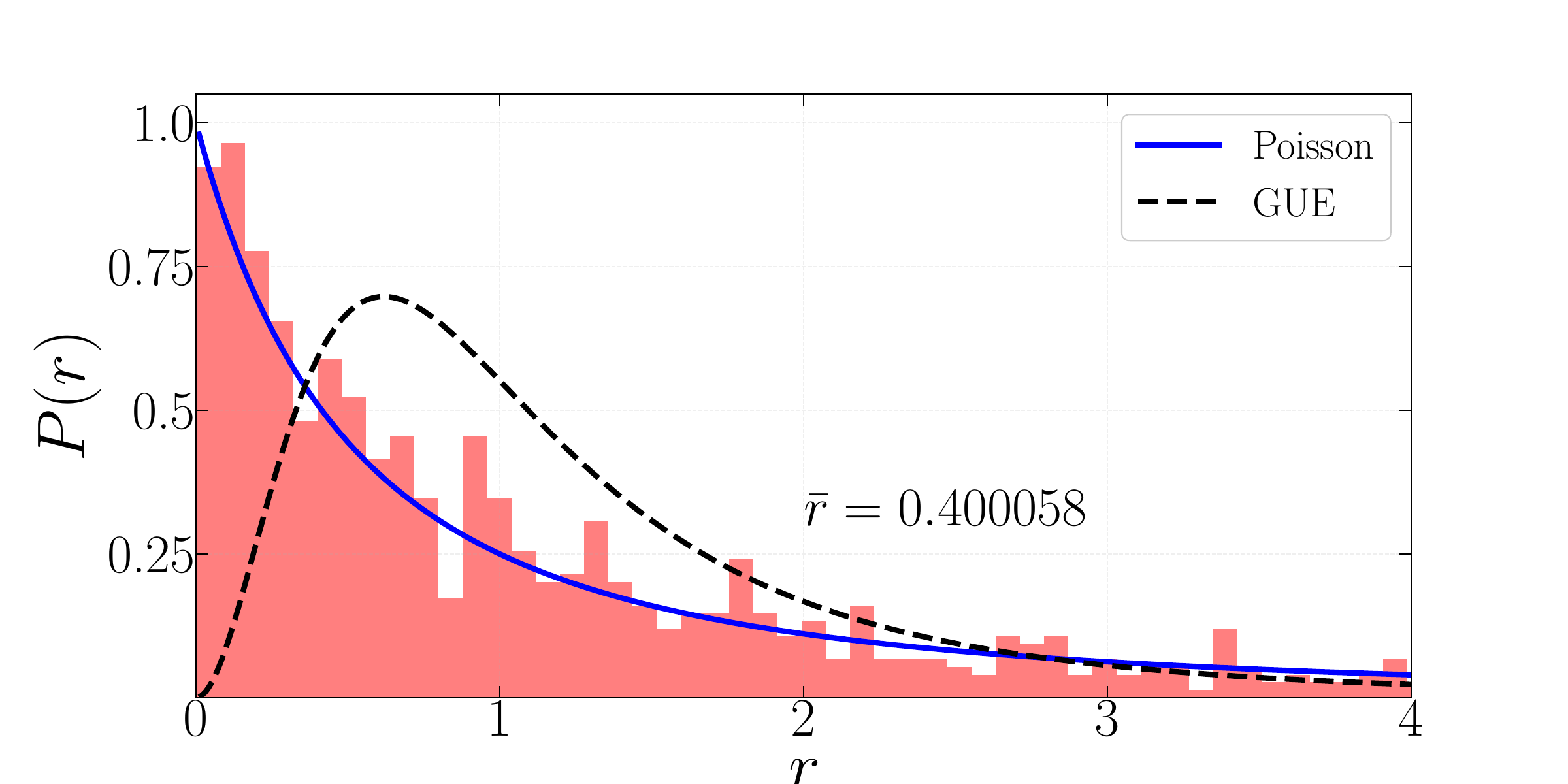}
\caption{Consecutive entanglement spectrum spacing ratio histogram of the reduced density matrix (RDM) eigenvalues of the ground state of the non-integrable Hamiltonian in eq.~\ref{HamDef2}. 
We also report the average spacing ratio. 
The Hamiltonian parameters used are: $J \!=\! -1.25, J' \!=\! 0.2$ and $h \! = \! 1$. 
The histogram has $125$ bins and, having a single ground state for this model, the number of sampled RDM eigenvalues is $1024$.  }
\label{InitialES-Int}
\end{figure}

\begin{figure}[b!]
	\centering
\includegraphics[width=\columnwidth]{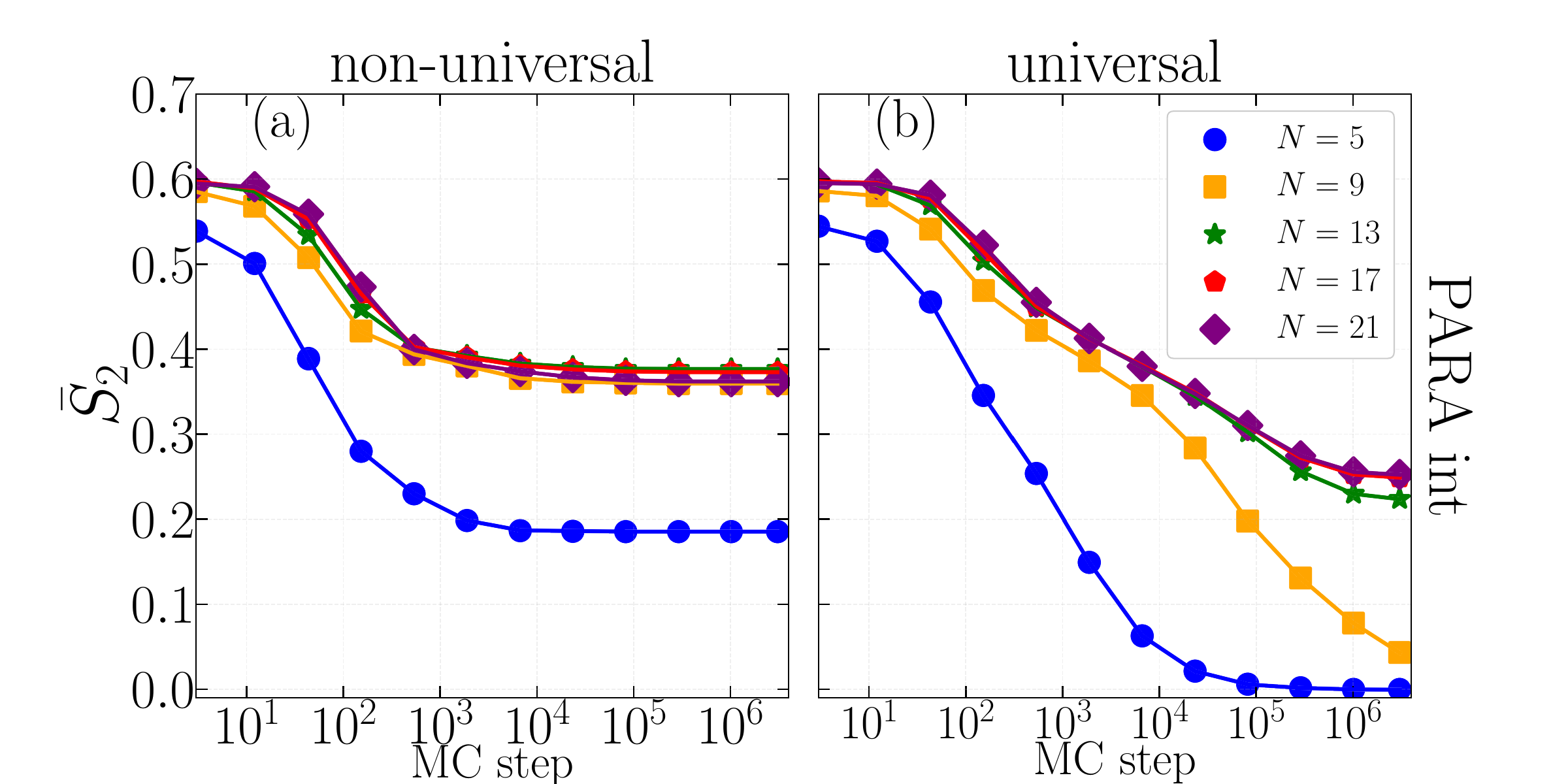}
\caption{Evolution of the entanglement entropy under the cooling algorithm, starting with the ground state of the non-integrable model with Hamiltonian in eq.~\eqref{HamDef2} in the paramagnetic phase with parameters $J \! = \! -1.25, J'\! = \! 0.2$ and $h \! = \! 1$. 
Data points represent the averaged half-chain R\'{e}nyi-2 entanglement entropy over $M = 96$ Metropolis MC trajectories. 
Both the non-universal (panel $(a)$) and universal (panel $(b)$) gate sets (see Table~\ref{table1}) are considered in the entanglement cooling procedure. }
\label{EntEvolution-Int}
\end{figure}

\begin{figure*}
	\centering
\includegraphics[width=18cm]{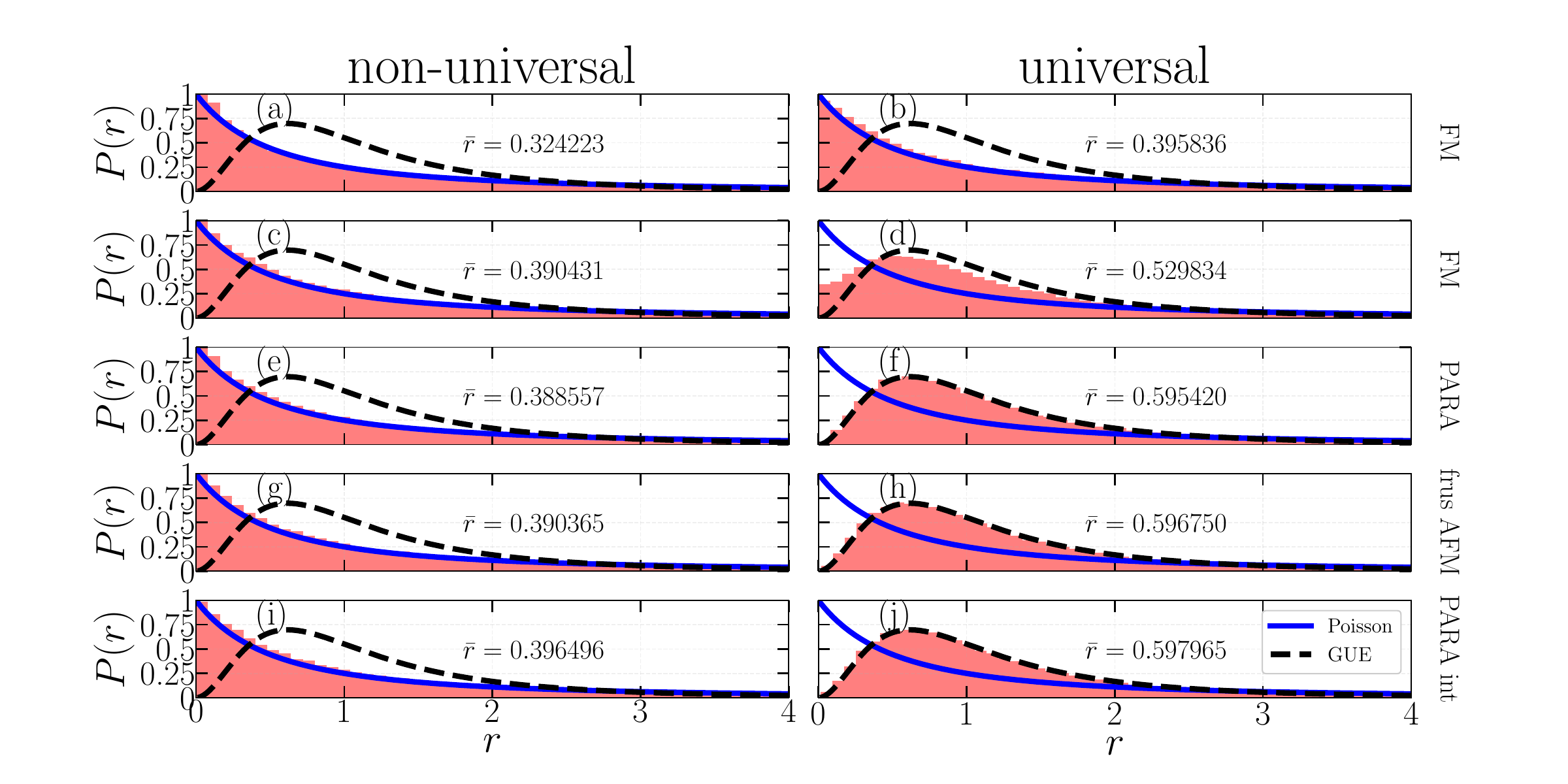}
\caption{Consecutive entanglement spectrum spacing ratio histogram of the reduced density matrix (RDM) eigenvalues for the final states at the end of the entanglement cooling algorithm for the integrable TFIM and its non-integrable extension, the ANNNI model in its paramagnetic phase. 
The states used for these averages over $M =96$ samples are those obtained in Fig.~\ref{EntEvolution} and \ref{EntEvolution-Int}. 
The histograms have been generated using $125$ bins. 
In the left panels, resulting from an algorithm employing only parity preserving gates, the final entanglement spectrum statistics is always Poissonian as it was at the beginning (see Fig.~\ref{InitialES} and \ref{InitialES-Int}). 
In the right panels, obtained using universal gates, the entanglement spectrum has always evolved toward a Wigner-Dyson distribution, except for the deep ferromagnetic case, characterized by very little initial local entanglement.}
\label{FinalES}
\end{figure*}

On the contrary, in the AFM phase, due to the simultaneous presence of an odd number of spins, periodic boundary conditions, and antiferromagnetic interactions, the system exhibits topological frustration.
In this phase (due to the absence of an energy gap) the quasi-adiabatic continuation cannot be straightforwardly applied. However, as shown by previous works~\cite{Giampaolo2019, Odavic2022}, in several respects the ground state can still be seen as a global state with multiparty entanglement on top of which, through a local deformation, an entanglement of local nature has been generated. 
Unlike the FM phase, the global state is no more represented by a GHZ state but by a sort of $W$-state, that is, in physical terms, a linear superposition of single kink states~\cite{Odavic2022}. 
This fact is crucial since, differently from the first kind of states,  part of the entanglement of the $W$-states is local, since they are the family of states with multipartite entanglement~\cite{Oliveira2006a, Oliveira2006b, Anfossi2008, Giampaolo2013, Giampaolo2014, Giampaolo2015, Hofmann2014, Girolamo2017, Gabbrielli2018, Haldar2020} that maximize the amount of bipartite entanglement after local measurement on one of its part~\cite{Dur2000, Coffman2000}. 
The presence of such bipartite entanglement, absent in the states coming from the FM phase, provides a path along which the entanglement cooling algorithm can act, reducing, but not removing, the amount of entanglement in the system. 
Moreover, for every single pair of spin, the amount of bipartite entanglement in a $W$-state reduces with the system size, hence explaining why the difference between the values of the initial and final plateaus in the dynamics of the $W$-state reduces as $N$ increases.

\subsubsection*{A Non-integrable Extension: the ANNNI Model}
\label{sec:nonint}

As we mentioned above, the TFIM is also known for being exactly solvable. 
While we do not expect this feature to influence the behavior of the entanglement under the cooling algorithm, to make sure that this is indeed the case, we consider now a non-integrable extension of the TFIM, namely the ANNNI model~\cite{Peschel1981, Selke1988, Chandra2007}:
\begin{equation}
H = J \sum\limits_{j =1}^{N} \sigma_{j}^{x} \sigma_{j+1}^{x} + J' \sum\limits_{j =1}^{N} \sigma_{j}^{x} \sigma_{j+2}^{x} - h \sum\limits_{j = 1}^{N} \sigma_{j}^{z},  
\label{HamDef2}
\end{equation}
where we assume periodic boundary conditions and focus on the ferromagnetic regime $J\!=\!-1.25$ with $h\!=\!1$. 
The next-to-nearest neighbor interaction we added compared to the traditional TFIM in  eq.~\eqref{HamDef} breaks the integrability. 
We fix this additional coupling to be antiferromagnetic $J'\!=\! 0.2$, to add some amount of frustration and correlation in the system, while keeping it subdominant with respect to the larger nearest neighbor coupling to ensure we do not cross a transition to a extensively frustrated phase. The zero-temperature phase diagram of this model has been worked out over the years in the literature~\cite{Chandra2007, Allen2001, Guaimaraes2002, Beccaria2007, Suzuki2013}. 
The relatively high magnetic field pushes the system into a paramagnetic phase. 
While the integrability of the clean TFIM allowed us to construct its ground states for relatively large chains, for this model we are not able to push the system size sufficiently: in Fig.~\ref{InitialES-Int} we report the level spacing statistics for the initial state entanglement spectrum, which looks Poissonian, although the limited statistics give rise to some spurious fluctuations.
In Fig.~\ref{EntEvolution-Int}, we present the results of the entanglement cooling algorithm using both available gate sets.
We observe similar results as in the paramagnetic phase of the integrable TFIM (see panels $(e)$ and $(f)$ in Fig.~\ref{EntEvolution}). Overall, the universal gate set is better at entanglement cooling compared to the non-universal one.
We decided to test the paramagnetic phase of the ANNNI model in order to have purely local entanglement for our algorithm to act upon, and we notice that starting with a higher value compared to the TFIM, in the end, the algorithm settles also on a higher plateau.

\subsubsection*{Entanglement Spectrum Statistics}
\label{sec:entspectrum}

As discussed before, the entanglement spectrum statistics of all initial states we considered follow a Poisson distribution (with deviations related to exact degeneracies), as is often the case with the ground states of the local Hamiltonian. 
In Fig.~\ref{FinalES} we report the average entanglement spectrum statistics obtained at the end of the cooling algorithm for all cases we considered so far (namely the final points for the larger system sizes in Fig.~\ref{EntEvolution} and \ref{EntEvolution-Int}). 

\begin{figure}
\centering	\includegraphics[width=\columnwidth]{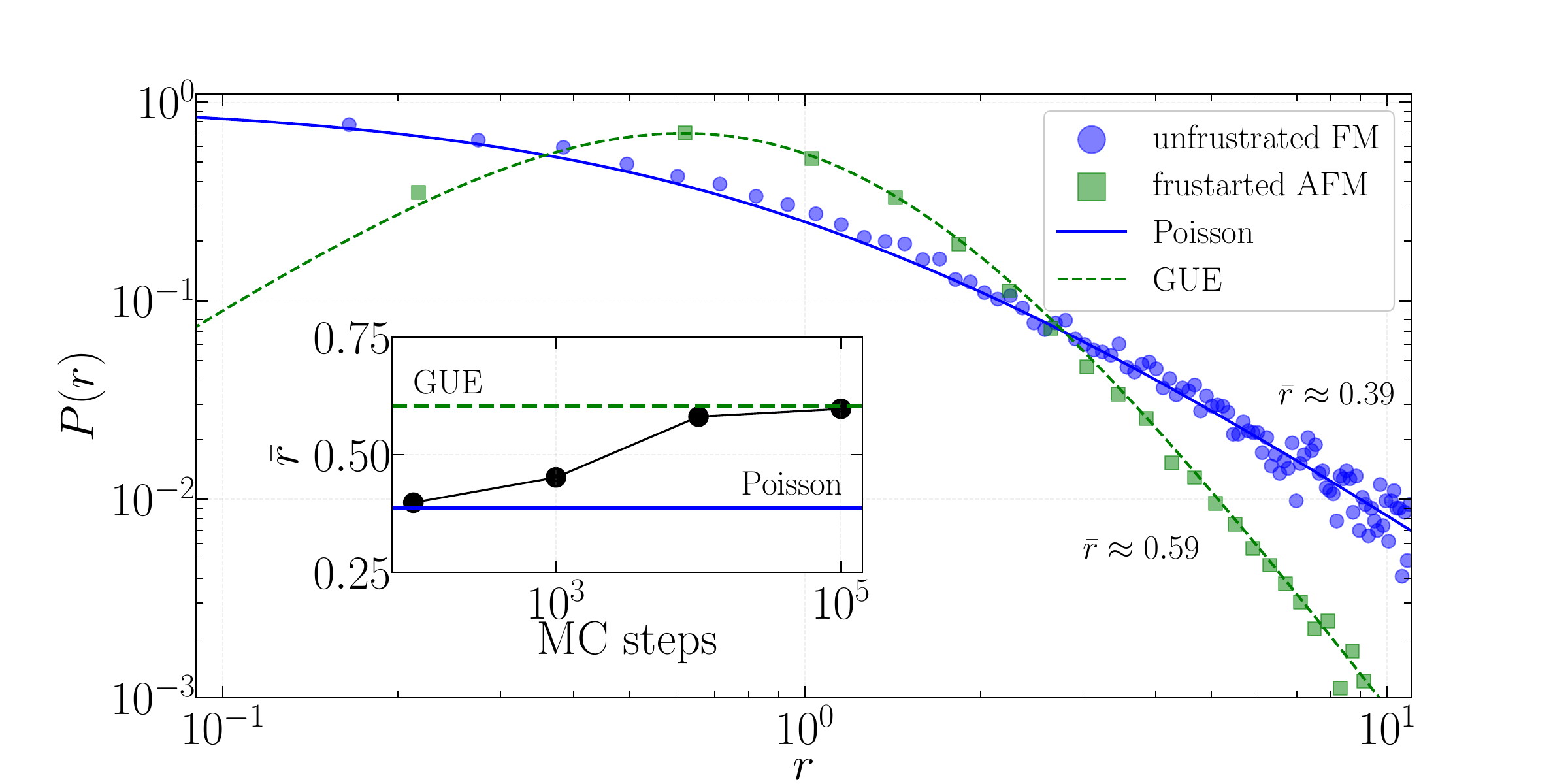}
\caption{ Entanglement spectrum eigenvalue differences of the unfrustrated deep ferromagnetic ($J/h = -2.5$) vs frustrated AFM ($J/h = 2.5$) state for system size $N = 17$ after the application of the cooling algorithm with universal gates for $10^{5}$ steps with $10^2$ realizations. 
The predictions from RMT are given with lines.  
Inset shows the average consecutive spacing ratio with variable Metropolis MC steps for the frustrated system of size $N = 17$. 	}
\label{EntComp}
\end{figure}

While using only the first gate set, the entanglement spectrum remains Poissonian, using universal gates we always observe an evolution toward exact Wigner-Dyson statistics in all cases, except when starting with a state with negligible local entanglement, such as in the deep ferromagnetic regime. 
Our interpretation is that, in the process of removing local entanglement, our algorithm also performs a sort of scrambling which results in the observed ``chaotic'' spectrum. 
But this can happen only if there is sufficient local entanglement to act upon, meaning that a sufficiently large number of moves are accepted before reaching the final plateaus. 
In Fig.\ref{EntComp} we compare the consecutive RDM level spacing statistics for the final states obtained using universal gates in the deep ferromagnetic case and the frustrated AFM one to highlight their difference and to show the evolution of the entanglement ratio during the cooling process as a function of the number of Montecarlo steps. 
We stopped the evolution after an equivalent amount of Metropolis MC step applications, and we choose this number to be $10^{5}$ steps, but the results are the same if we evolve the state even further.

\begin{figure}[t!]
\centering
\includegraphics[width=\columnwidth]{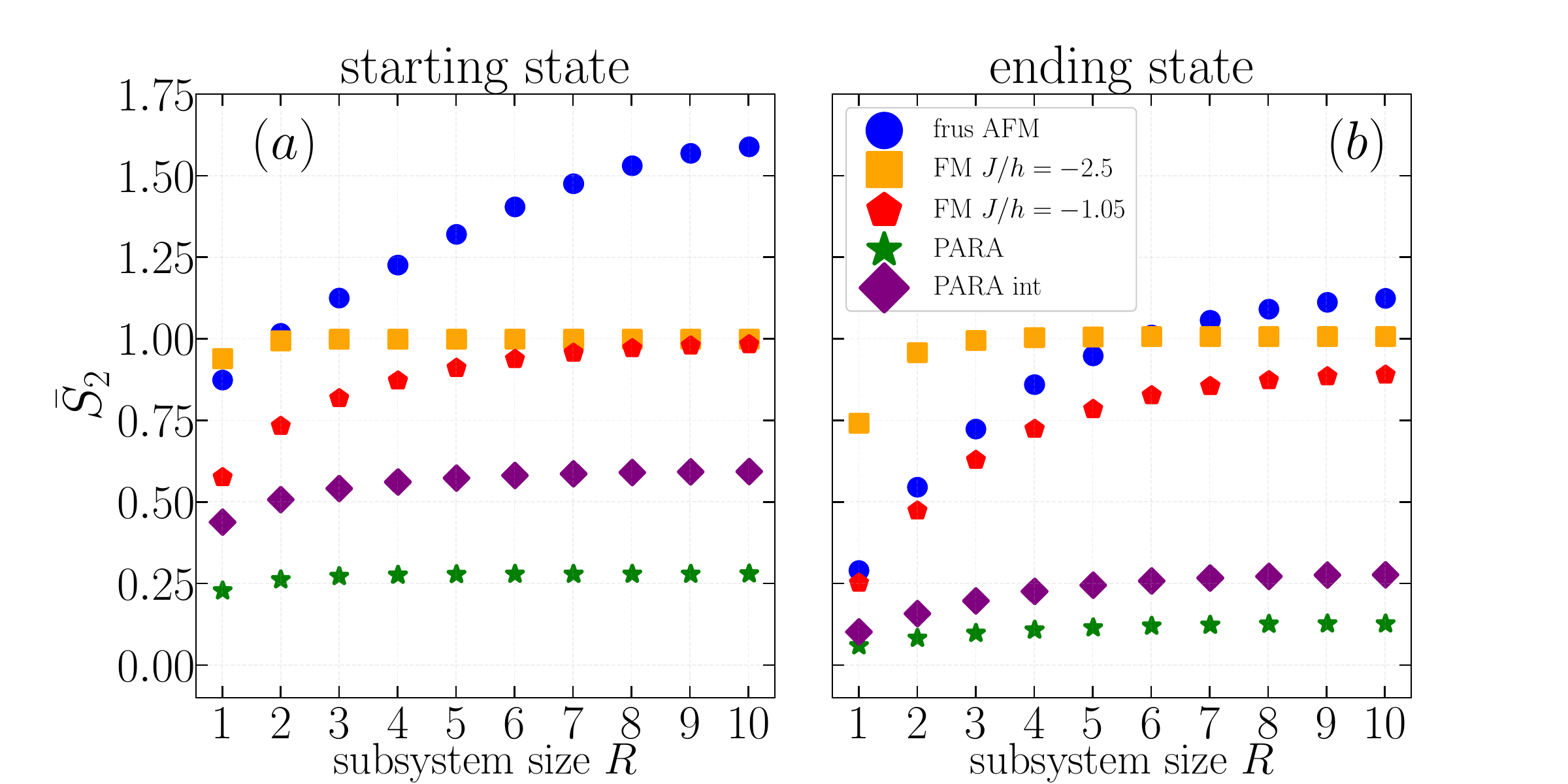}
\caption{Averaged half-chain R\'{e}nyi-2 entropies as a function of the subsystem size. 
Left $(a)$ panel: Entanglement dependence on the subsystem size for the different initial states used in Fig.~\ref{EntEvolution} and \ref{EntEvolution-Int}. 
Right $(b)$ panel: Same for the states at the end of the cooling algorithm, that is, for the same states used in Fig.~\ref{FinalES}.
Although for frustrated cases there is some residual entanglement growing with the subsystem size, it is clear that none of these entropies satisfy an area law.}
\label{AreaLaw}
\end{figure}

We remark that, up to now, the appearance of a Wigner-Dyson entanglement spectrum has always been associated with the presence of a volume law for the corresponding entanglement entropy.
In Fig.~\ref{AreaLaw} we present the values of the R\'{e}nyi-2 entropy as a function of the subsystem length for the initial and final states in our analysis. 
As expected, the initial ground states of our Hamiltonians in their various phases always satisfy an area law, with a relative violation for the frustrated case where the entanglement can grow with the system size, while remaining upper-bounded, as explained in Ref.~\cite{Giampaolo2019}. 
Similarly, the final states obtained at the end of the algorithm never approach a volume law and are compatible with an exact area law. 
To the best of our knowledge, these results are unprecedented and deserve additional analysis in the future, since they imply that it is possible to construct local models with area law whose entanglement spectrum follows a Wigner-Dyson distribution, although they have never been found so far.

\section{Conclusion \label{sec:conclusion}}

In our work, we tested the evolution of entanglement induced by applying a cooling algorithm to different states, all but one being the ground states of Hamiltonians in the same class of integrable models, but in different macroscopic phases. 
For completeness, to check the independence of our results from integrability, we also considered a non-integrable extension.
Each of the chosen states is characterized by an initial entanglement spectrum following the same Poissonian distribution.
Quite surprisingly, states in different phases evolve differently, depending on the macroscopic phase and on the different sets of gates employed.
In the paramagnetic phase, the ground state can be obtained by adiabatic continuation from a fully factorized, classical one. 
Thus, it contains only local correlations whose entanglement the cooling algorithm can progressively remove when a complete (universal) set of local unitaries are employed.
In the ferromagnetic phase, in agreement with the quasi-adiabatic continuation approach, the ground states can be seen as locally deformed GHZ states~\cite{Hamma2016, Hastings2005}. 
Hence, after the cooling algorithm has destroyed the local entanglement, the residual global entanglement (characteristic of a GHZ state) is resilient to such a process.
Its destruction would require a series of moves that are accepted even if they are increasing the entanglement until they overcome the barrier separating a GHZ state from a separable one, but such an event is exponentially suppressed with the system size.

Finally, in the presence of topological frustration, we are not in the range of validity of the quasi-adiabatic continuation. 
However, it was proved~\cite{Giampaolo2019, Odavic2022} that the ground state can still be seen as a locally deformed W-state.
The $W$-states, as well as the GHZ states, are states that, while maximizing the multipartite entanglement, also show a large bipartite entanglement, unlike the latter states. 
Its presence, being also somewhat non-local, provides an amount of entanglement that the cooling algorithm can destroy, but less efficiently as the system size increases.

Overall, our results show a much more varied picture than what one might have expected. 
The reason why it remained unnoticed until now is mostly due to the fact that previous analyses have started from randomly generated states in which geometrical and topological properties are absent.
Indeed, the three phases we considered present entanglements of three different types which are not reflected by the entanglement spectrum statistics: purely local, GHZ-like, and W-state-like, with the topologically frustrated chain exhibiting all of them. 
Indeed, our data suggest that the evolution induced by the cooling algorithm is affected by the geometrical and topological properties of the ground state manifold of the initial state, as well as the presence, and the kind, of multipartite entanglement. 
Moreover, we observe that the choice of gate set affects significantly the amount of entanglement that the algorithm is able to destroy and thus the value of the final plateaus.

In fact, in all cases, we observe that, contrary to what one could have expected, the entanglement cooling algorithm is never able to completely destroy all local entanglement. 
It rather settles on plateaus in which some sort of meta-stable equilibrium is reached between the amount of entanglement that is reduced at a given step and the one that is added by a move accepted because of the finite temperature in the annealing process.

A striking, completely unexpected, behavior, which we believe is connected with the nature of the plateaus we just discussed, is that when a sufficient amount of local entanglement is present in the initial state and eventually destroyed by the algorithm.
In these cases, our results show that the final state features a Wigner-Dyson distribution of the level statistics of its entanglement spectrum. 
Effectively, while cooling, the algorithm has also scrambled the entanglement spectrum and increased its complexity. 
As Wigner-Dyson statistics has been connected with higher robustness for it, at this point the algorithm remains stuck in a plateau, unable to destroy additional entanglement effectively. 
Note that, since this statistic has been observed before only in states obeying a volume law for the entanglement entropy, our findings disprove the expectation that the two are connected. 
It would be interesting to understand what are the conditions for which local Hamiltonians can generate area law ground states with the Wigner-Dyson entanglement spectrum, but this is a subject for future work.

Before concluding, we underline that our work is relevant from a methodological point of view as well. 
In particular,  we use GPU-enhanced calculations to accelerate and scale the entanglement cooling protocol. 
The numerical codes, data, and plotting scripts employed for our cooling algorithm are provided in Refs.~\cite{zenodo,github}.

\begin{acknowledgments}
S.M.G., F.F., and G.T. acknowledge support from the QuantiXLie Center of Excellence, a project co–financed by the Croatian Government and European Union through the European Regional Development Fund – the Competitiveness and Cohesion (Grant KK.01.1.1.01.0004). 
F.F. and S.M.G. also acknowledge support from the Croatian Science Foundation (HrZZ) Projects No. IP–2019–4–3321.
J.O., N. M. and D.D. recognize the support of the Croatian Science Foundation under grant number HRZZ-UIP-2020-02-4559 and the European Regional Development Fund under the grant KK.01.1.1.01.009 - DATACROSS. 
The authors gratefully acknowledge the HPC RIVR consortium (www.hpc-rivr.si) and EuroHPC JU (eurohpc-ju.europa.eu) for funding this research by providing computing resources of the HPC system Vega at the Institute of Information Science (www.izum.si).
Part of this research was performed using the advanced computing service - supercomputer Supek - provided by the University of Zagreb University Computing Centre - SRCE.
\end{acknowledgments}

\appendix

\section{\label{sec:Appendix1} Computational details and speed-up of execution}

In the cooling algorithm, the computation of the Rényi-2 entropy (see Sect. II) is the major computational barrier. 
Although diagonalization is replaced by matrix-matrix multiplication to calculate entropy, it still remains the main bottleneck, accounting for most of the execution time. 
Since the size of the matrix and the number of matrix-matrix multiplications (i.e. the number of Monte Carlo steps) cannot be reduced, the algorithm can be speed-up by executing the matrix multiplication on graphical processing units (GPU). 
GPUs are best suited for matrix operations due to their massively parallel (many cores) architectures, delivering up to an order of magnitude better performance compared to (multithreaded) CPUs. 
The biggest disadvantage is that GPUs cannot perform well if the matrices are too small. 
In this case, a slowdown can even be observed compared to CPU. 
In our experiments, we display systems with $N = 21$ spins using matrices of size $2048 \times 1024$, which are too small to achieve good utilization of GPU resources. 
The solution is to pack multiple matrix multiplications into a single and large matrix-matrix multiplication operation called batched operation to achieve high performance. 
For this particular research, we have proposed a different approach based on multiple concurrent tasks instead of a single batched operation. 
The detailed implementation, speed-up achieved and performance discussion can be found in~\cite{Mijic2022}, while we give only a brief description below.

The mutual independence of the Metropolis MC simulations provides a trivial parallelization, as no synchronization or communication is required, making the problem embarrassingly parallel. 
The Message Passing Interface (MPI) is used to associate each trajectory with a single MPI process, while multiple processes can be run concurrently on the same device using the NVIDIA Multi-Processing Service (MPS). 
This approach allows spatial sharing of device resources with the disadvantage of slowing down the computation of a single Metropolis MC trajectory (i.e. matrix multiplication), but this has the advantage of allowing multiple simulations to be computed in parallel, effectively overlapping the computations from different trajectories. 
Analysis has shown~\cite{Mijic2022} that overlapping has a much better effect on overall performance than slowing down the single Metropolis MC trajectory. 
For the largest use case with a system size of N = 21 spins and 96 MC trajectories, we found that the best configuration is 12 procedures (MPI processes) per NVIDIA A100 GPU, which took about 116 hours on 8 GPUs. 
Overall, we achieved a speed-up of two orders of magnitude on a single Metropolis Monte Carlo (MC) trajectory compared to the standard CPU
implementation.
\begin{figure*}[t!]
    \centering
    \includegraphics[width=18cm]{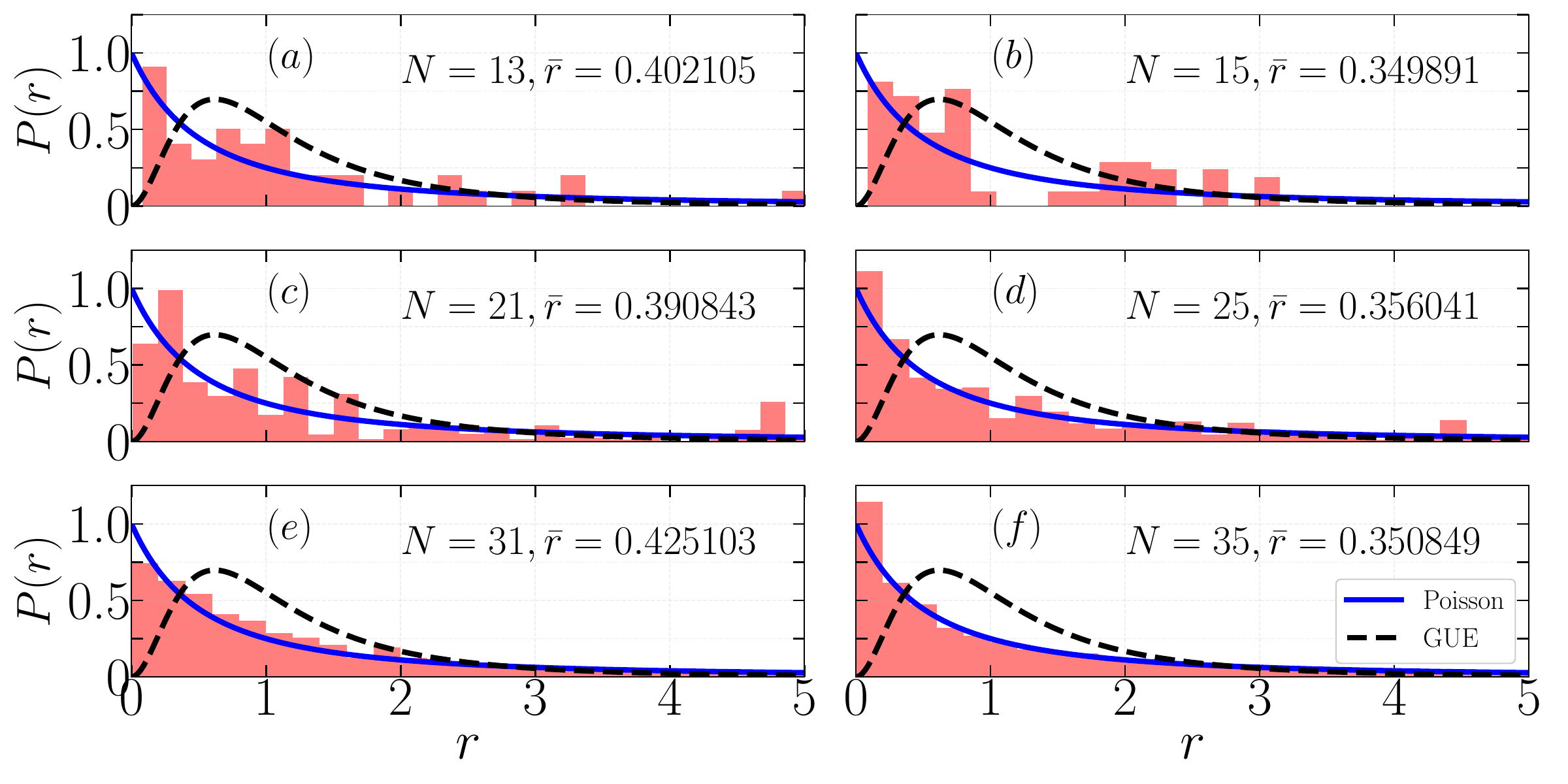}
    \caption{ Consecutive entanglement spectrum spacing ratio histogram of the RDM eigenvalues for different systems size. 
    The chain sizes are $N = 13,15,21,25,31,35$; while the half-chain sizes of the RDM are $\lfloor N/2 \rfloor = 6,7,10,12,15,17$. 
    We only focus on the range $0<r<10$ where we consider 50 bins to make the histogram. 
    Inside the plot, we provide the averaged consecutive spacing ratio number we compute from the given spectrum. Parameters for this case are $h = 1.0$ and $J = 2.5$ which is the antiferromagnetic frustrated regime. 
    This is single-shot data ($M = 1$), and no ensemble averaging is performed.}   
    \label{ESvsSizes-50Bins}
\end{figure*}

\section{\label{sec:Appendix2}Entanglement spectrum statistics details}

\textit{Consecutive level spacing ratios}: In the manuscript, we focus on the consecutive level spacing ratio probability density function $P (r)$. 
This is a suitable quantity of interest in the
case of quantum many-body applications where statistics quality is heavily bound by the Hilbert space dimension~\cite{Oganesyan2007, Atas2013, Odavic2021}.  In particular, using an ascending (from smallest to the biggest) ordered set of eigenvalues $\{ \lambda_{k} \}$  of the reduced density matrix, we compute the consecutive ratios as 
\begin{equation}
    r_{k} = \frac{\lambda_{k + 1} - \lambda_{k}}{\lambda_{k} - \lambda_{k - 1} }, \quad k = 2, 3, 4, ..., 2^{ \lfloor N/2 \rfloor} - 1.\label{EQ1}
\end{equation}
Taking the values of the generated list of $\{ r_{1}, r_{2}, ...\}$ we plot a normalized histogram, where we disregard any spurious (and rare) values that are larger than $r_{j} > 10.0$ for proper normalization and binning of the histogram.

\textit{Averaged consecutive spacing ratio:} Additionally, we consider the averaged consecutive spacing ratio defined as 
\begin{equation}
\bar{r} = \Bigg\langle \Bigg\langle \frac{\min (s_{k,j},s_{k+1,j})}{\max(s_{k,j},s_{k+1,j})} \Bigg\rangle \Bigg\rangle_{2^{ \lfloor N/2 \rfloor} -2,M}, \label{EQ2}
\end{equation}
where we define the spacings as $s_{k,j} = \lambda_{k+1,j} - \lambda_{k,j}$, and the index $j = 1, 2, ..., M$ refers to the different reduced density matrices considered (ensemble). 

From Random Matrix Theory (RMT) it is known that, in the Poisson case, the ratio follows the probability distribution $P_{\rm Poisson} (r) = (1 + r)^{-2}$.
On the contrary, for the Wigner-Dyson distribution, we have $P_{\rm WD} = Z_{\beta}^{-1} (r + r^2)^{\beta} (1 + r + r^{2})^{-1 - 3/2 \beta}$ with $\beta = 2$ (Gaussian Unitary Ensamble - GUE) and $Z_{\beta = 2} = 4 \pi/81 \sqrt{3}$. 
The predictions for the averages are thus $\bar{r}_{\rm Poisson} = 2 \ln{2} - 1 \approx 0.386$ and $\bar{r}_{\rm WD} = 2 \sqrt{3}/\pi - 1/2 \approx 0.602$~\cite{Atas2013}.

\textit{Advantages for the integrable TFIM:} The integrable TFIM can be mapped to free-fermions using the Jordan-Wigner transformation~\cite{FranchiniBook} and solved explicitly ~\cite{McCoy}. 
The half-chain reduced density matrix (RDM) $\rho_{A}$ of the ground state for this model can then be computed using the prescription detailed in Ref.~\cite{VLRK}. 
The domain $A$, due to translational invariance of the spin chain, comprises any $ \lfloor N/2 \rfloor $ (half-chain) contiguous spins. 
Note that with this approach, we are able to push our numerical considerations much beyond the Exact Diagonalization (ED) efficiency~\cite{Sandvik2010}. 
Additionally, using \textit{mpmath}~\cite{mpmath} Python library we boost the precision of the fermionic approach, as double precision does not suffice for this kind of analysis. 
For the largest example, we shall consider, i.e. $N = 35$ and half-chain of size of the RDM of $17$ the number of eigenvalues of the RDM is $131072$. 
In Fig.~\ref{ESvsSizes-50Bins} we show the results for the entanglement spectrum statistics for a single ($M= 1$) realization and different chain lengths, to show how important is to reach adequate system sizes to obtain reliable statistics.

\end{document}